\documentclass[twocolumn,showpacs,aps,prb]{revtex4}% Physical Review B
\usepackage{graphicx} % Include figure files
\usepackage{dcolumn} % Align table columns on decimal point
\usepackage{bm} % bold math
\usepackage{amsmath}

\begin{document}
\preprint{Physical Review B}
%\title{Electronic transport properties of carbon nanotubes with Kondo impurities}
%\title{Kondo effects on the transport in carbon nanotubes with magnetic impurities}
\title{Conductance and thermoelectric power in carbon nanotubes with magnetic impurities}
\author{Fufang Xu}
\affiliation{%
Department of Physics, Tsinghua University, Beijing 100084, China}
\author{Jia-Lin Zhu}
\email[Electronic address: ]{zjl-dmp@tsinghua.edu.cn}
\affiliation{%
Department of Physics, Tsinghua University, Beijing 100084, China}
\date{\today}% It is always \today, today,

\begin{abstract}
By introducing a normal self-energy to incorporate the effects of magnetic impurities, Kondo
effects in single-walled metallic carbon nanotubes are investigated within the Anderson model and
Landauer formula. Magnetic impurities induce Kondo resonance and bring a valley of conductance
function near the Fermi level. The conductance of the nanotube increases with the temperature in
the low temperature range. Thermoelectric power induced by magnetic impurities is gotten from Mott
relation, and the calculations indicate its dependence on the temperature could interpret the
experiment well.
\end{abstract}

\pacs{73.23.-b, 73.63.Fg, 75.20.Hr} % PACS, the Physics and Astronomy

\maketitle
\section{Introduction}
Research on the carbon nanotubes has attracted much interest due to their unique properties and
great application potentials. A single-walled carbon nanotube (SWNT) can be metallic or
semiconducting depending on its radius and chirality.\cite{Saito} The properties of carbon
nanotubes can be greatly modified by defects or impurities,\cite{Song} which may be introduced in
the production of nanotubes or be doped artificially. The role of magnetic atoms or clusters in
the carbon nanotubes is interesting and may have significant effects on the electronic transport
properties. The earlier thermoelectric power (TEP) measurements on single walled carbon nanotube
mats suggest that the observed giant TEP comes from the contributions of Kondo state induced by
magnetic transition-metal catalysts.\cite{Grigorian} In addition, low-temperature scanning
tunnelling microscopy (STM) spectroscopy is probed for isolated nanometer ferromagnetic cobalt
clusters on surface of metallic single-walled carbon nanotubes. The results clearly indicates the
appearance of Kondo resonance.\cite{odom}

%%Kondo的历史和现状 说明其他kondo缺少一个有效的过渡
Magnetic impurities in non-magnetic metallic host, which are called Kondo impurities, have been
explored in bulk systems for many years. The most attracting features are that they bring a series
of {\it anomalies} on the physical properties of the host, including conductance, specific heat
and thermoelectric power (TEP) and so on.\cite{hewson} They are generally called Kondo effects.
Recently, the Kondo effects in mesoscopic systems, such as quantum dots and carbon
nanotubes,\cite{surfkondo,qdkondo,cntkondo} arouse many attentions for their unique properties. In
a semiconductor quantum dot or a carbon nanotube quantum dot, the conductance shows different
behavior depending on whether the number of electrons confined in the dot is even or odd. By
introducing magnetic impurities into a carbon nanotube, an one-dimensional Kondo system is formed
\cite{Costa,Shenoy} and may lead to some important applications. The schemes by employing this
magnetic impurities/SWNT systems to generate quantum entanglement states for quantum information
processing, a potential field in the future, have been proposed. \cite{entangle, entangle2}

%%%cite main Kondo papers on CNQDs
%%Kondo,尤其是@CNT的计算,引出磁性掺杂对输运性质的影响以及我们的计算
A theoretical study on magnetic clusters in carbon nanotubes has given STM spectra and the
dependence of Kondo temperature on the size of magnetic clusters.\cite{fiete} Besides that, the
density of states (DOS) for metallic carbon nanotubes with a magnetic impurity is investigated by
utilizing the perturbation theory, and give results which agree well with the
experiments.\cite{wei1,wei2} However, the transport properties of a metallic SWNT with magnetic
impurities have not been investigated systematically so far. In this paper, Anderson model is used
to describe the system and the transport properties are studied within the Landauer formula. The
conductance and the TEP at finite temperature are explored. The relation between the temperature
and the Kondo effect is found to interpret the experiment well.

\section{Model and Formula}
%%%Anderson  model
The magnetic impurities/SWNT system can be assumed that it gets the contribution from the sum of
the isolated impurities, if the impurity density is dilute and the interaction among them can be
neglected. Then it can be described by a single-orbital Anderson model.\cite{anderson} The
Hamiltonian of the whole system can be written as:
\begin{eqnarray}
 H&=&\sum_{k,\sigma}\varepsilon_kC^+_{k,\sigma}C_{k,\sigma}+\sum_{\sigma}\varepsilon_dC^
 +_{d,\sigma}C_{d,\sigma}\nonumber\\
 &&+\sum_{k,\sigma}(V_kC^+_{k,\sigma}C_{d,\sigma}+V^*_kC^+_{d,\sigma}C_{k,\sigma})
 +Un_{d,\uparrow}n_{d,\downarrow}\nonumber\\
\end{eqnarray}%
where the first term describes the conduction electrons of nanotubes in a frame of tight-binding
model. $C^+_{k,\sigma}$ and $C_{k,\sigma}$ are creation and annihilation operators for Bloch
states of wave vector k and spin component $\sigma$, corresponding to energy $\varepsilon_k$. In
the second term, $\varepsilon_d$ is the energy of the localized d level of the impurity, and
$C^+_{d,\sigma}$ and $C_{d,\sigma}$ are creation and annihilation operators for an electron in
this state. Here we ignore the orbital degeneracy of the d level and treat it as a state with spin
$\uparrow$ and spin $\downarrow$ degeneracy only. The third term describes overlap of localized
level with the wavefunction of conduction electrons in carbon nanotubes, and $V_k$ is the
hybridization matrix element. For the last term, $U$ is the onsite Coulomb repulsive interaction
between electrons in localized state, and $n_{d,\sigma}=C^+_{d,\sigma}C_{d,\sigma}$.

%%% EOM get Green function
The route to work on this Hamiltonian can be based on equations-of-motion method for the
double-particle imaginary-time Green function\cite{wei1}. Applying this method to the above
Hamiltonian, we get
\begin{eqnarray}
G_s(\varepsilon)&=& G^0_s(\varepsilon)+G^0_s(\varepsilon)\left(\sum_kV_kG^d(\varepsilon)V_k^*\right)G^0_s(\varepsilon)\\
G_d(\varepsilon)&=&
\left(\varepsilon-\varepsilon_d-\sum_kV_k\frac{1}{\varepsilon-\varepsilon_k}V^*_k-\Sigma_d\right)^{-1}
\end{eqnarray}

Comparing the $G_s$ with perturbation expansion $G=G_0+G_0 \Sigma'
G_0$, we get the self-energy of conduction electrons,
$\Sigma'_s=\sum_k V_kG_dV_k^*$.
%Once we get $\Sigma_s$,We can compute the conductance and TEP within
%the Landauer formula.
%%Gd, Self-energy.
Therefore, how to get $G^d$ is the key to implement calculation. A rapidly convergent perturbation
method was firstly proposed and developed in a series papers by Yamada and
Yosida.\cite{yamada,yosida} Afterwards, Zlati\'{c} and Horvati\'{c} extended this method to
calculate the DOS of localized electrons for the asymmetric non-degenerate Anderson
model.\cite{horvatic,zlatic} It has been verified that the DOS calculated according to the above
method of Zlati\'{c} et al. agrees well with the Bethe ansatz results in the dilute magnetic alloy
region.\cite{horvatic2} We apply their method to calculate the self-energy at finite temperature
up to the second-order correction.

The retarded second-order self-energy is obtained as
\begin{eqnarray}
\Sigma^R_d(\omega)&=&\frac{\Delta}{2}u^2\bigg\{
\int^{+\infty}_{-\infty} \textrm{tanh}(\frac{\beta \Delta
\varepsilon}{2})
\textrm{Im}[\tilde{G}^R_0(\varepsilon)]\tilde{\chi_0}(\frac{\omega}{
\Delta}-\varepsilon)d\varepsilon \nonumber\\%
&&+P\int^{+\infty}_{-\infty} \textrm{coth}(\frac{\beta
\Delta\varepsilon}{2})
\tilde{G}^R_0(\frac{\omega}{\Delta}-\varepsilon)\textrm{Im}[\tilde{\chi_0}(\varepsilon]d\varepsilon\bigg\} \nonumber\\
\end{eqnarray}%
where
\begin{eqnarray*}
\tilde{G}^R_0(\varepsilon)&=&(\varepsilon-\frac{E_d}{\Delta}+i)^{-1},\\
\tilde{\chi}_0(\varepsilon)&=&\frac{-1}{\varepsilon(\varepsilon+2i)}\Bigg\{\Psi\left[\frac{1}{2}
+\frac{\beta\Delta}{2\pi}(1+i \frac {E_d}{\Delta}-i\varepsilon)\right]\\%
& &+\Psi\left[\frac{1}{2} +\frac{\beta\Delta}{2\pi}(1-i \frac
{E_d}{\Delta}-i\varepsilon)\right]\\%
& &-\Psi\left[\frac{1}{2} +\frac{\beta\Delta}{2\pi}(1+i \frac
{E_d}{\Delta})\right]\\%
& &-\Psi\left[\frac{1}{2} +\frac{\beta\Delta}{2\pi}(1-i \frac
{E_d}{\Delta})\right]\Bigg\}.
\end{eqnarray*}%
Three parameters has been taken to describe the model:
$u=\frac{U}{\pi\Delta}$, $E_d=\varepsilon_d+\frac{1}{2}\langle
n_d\rangle$, and $\Delta=\frac{\pi}{2}\rho(0)V^2$. $\rho(0)$ the DOS
near the $E_F$ for the nanotube.

The self-energy $\Sigma_d(\omega)$ and the Green function $G^d(\omega)$ depend on temperature both
explicitly, through $\beta$, and implicitly, through the temperature dependence of $E_d$. So for
the case of finite temperature, the iteration is needed to reach the self-consistent results.

%%At finte to iteration
\begin{equation}
n_d=-\frac{2}{\pi}\int^\infty_{-\infty}d\varepsilon
f(\varepsilon/k_BT)\mbox{Im}G^R_{d\sigma}(\varepsilon)
\end{equation}

\begin{equation}
\begin{split}
G^R_{d\sigma}(\varepsilon)=\{\varepsilon-\Delta
[\frac{E^0_d}{\Delta}+u\,\mbox{tan}^{-1}\frac{E^0_d}{\Delta}\frac{1}{2}\pi
u(1-n_d)\\
+\Sigma^R_{d\sigma}(w,T,u,\frac{E^0_d}{\Delta})/\Delta]+i\Delta
\}^{-1}.
\end{split}
\end{equation}%
Here $E^0_d$ is taken as initial intput, the initial $n_d^0$ can be
judged from solve the transcendental equation.

\begin{equation}
\begin{split}
\mbox{cot}\left(\frac{\pi}{2}n^0_d\right)+\frac{\pi}{2}u(1-n^0_d)
=(E^0_d/\Delta)+\\u\,\mbox{tan}^{-1}(E^0_d/\Delta)
+\Sigma^R_{d,\sigma}(\varepsilon=0,T=0,E^0_d/\Delta)/\Delta.
\end{split}
\end{equation}

In order to investigate the transport properties, we take the whole system as a two-terminal
device. A long (10, 10) armchair nanotubes with magnetic impurities is investigated, and two
semi-infinite (10, 10) armchair nanotubes connected the are taken as ideal leads.

%%%%%%%%%%%%%%%%%%%%%%%%%%%%%
\section{Results and Discussion}
\subsection{Resistivity}
The transport through mesoscopic system can be calculated within the
Landauer formula.\cite{Datta}
\begin{eqnarray}
G(E,T)&=&G_0\Gamma^LG^r\Gamma^RG^a
\end{eqnarray}%
where $G_0=2e^2 /h$ is the conductance quanta. $\Gamma_{L(R)}$ is the coupling matrix between the
left (right) lead and the conductor. $G^r$ is the retarded Green's function which can be written
as
\begin{eqnarray}
G^r=\frac{1}{E-H^c_0-\Sigma^*_m-\Sigma^L-\Sigma^R}
\end{eqnarray}%
Here $H^c_0$ is the Hamiltonian matrix of the conductor that represents the interaction between
the atoms in the carbon nanotubes, and $\Sigma_{L(R)}$ is the self-energy function that describes
the effect of the left (right) lead, computed by using the surface Green's function matching
theory.\cite{Gfmatch} $\Sigma^*_m$ is the normal self-energy function that incorporate the effects
of magnetic impurities, and relates to the self-energy $\Sigma'$ through
\begin{eqnarray}
\Sigma^*_m=\frac{\Sigma'_s}{1+G^s_0\Sigma'_s}.
\end{eqnarray}%
As stated above, $\Sigma'_s$ can be gotten by comparing with the results with the perturbation
expansion of Green's function.

%%%%%%%%%%%%%%%%%%%%%
\begin{figure}
\includegraphics[angle=0,width=0.45\textwidth] {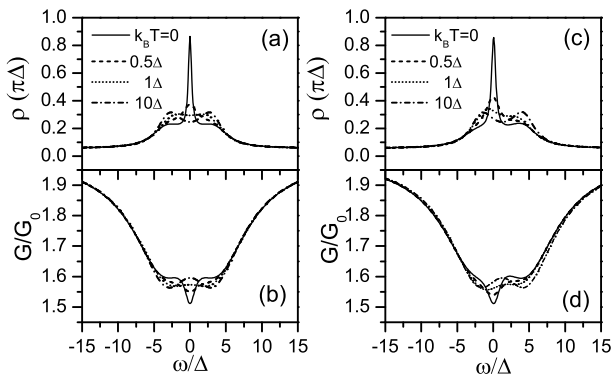}
\caption{\label{FIG:resis} Density of states and conductivity of a
(10,10) nanotube with magnetic impurities for (a, b) Symmetric case,
$u=2.0$ and $Ed=0$, (c, d) Asymmetric case, $u=2.0$ and
$Ed=0.5\Delta$ at different temperature}.
\end{figure}

Fig. \ref{FIG:resis}(a) and (b) give the DOS and the conductance function for the symmetric case
($E_d=0$) with u=2.0. At $T=10^{-5}K$, a triply peaked structure appears. A narrow resonant peak
appears at the $E_F$, and is called Kondo resonance. It is induced by the flip scattering due to
magnetic impurities. The two broad peaks, one below and one above the $E_F$, are two localized
peaks caused by effective local impurity energy level. Their separation is approximately $U$.
Correspondingly, the conduction function shows valley, as shown in (b). The narrow valley at the
$E_F$ is just result from the Kondo resonance. With the increase of temperature, the Kondo
resonance and the conductance valley fades out gradually. The two broad peaks are kept and  their
position is given in Hartree-Fock approximation. The results for an asymmetric case with
$Ed=0.5\Delta$ are given in Fig. \ref{FIG:resis}(c) and (d). The two broad peaks are not symmetric
with respect to $E_F$ any longer, and the position of Kondo resonance is little deviated from the
$E_F$. The conduction function possess the same character. In experiment, the DOS can be detected
by the differential conductance through STM measurement.
%%%%%%%%%%(n,n)

$\Delta$ parameterizes the solution of Anderson model and is proportional to the radius of
nanotube. It is determined by the hybridization between conduction electron and the impurity
electron, $V$, and the density of states of nanotubes near the $E_F$ ,$\rho(0)$. For a (n,n)
armchair nanotube, $\rho(0)$ is inversely proportional to the radius. Then $\Delta$ is
proportional to n. Then the results will be suppressed for the nanotube with bigger radius. For a
certain type magnetic impurity, Coulomb interaction $U$ is not related to the radius of nanotube.
So parameter $u=U/\pi\Delta$ is inversely proportional to the radius.

%%%%%%%%%%%%%%%%%%%%%%%
The conductivity at finite temperature can be got as:
\begin{eqnarray}
G(T)=G_0\int\limits_{-\infty}^{\infty}\left(-\frac{\partial
f}{\partial E} \right)T(E)\mbox{d}E
%G(T)=\pi\Delta\rho_d(\omega)
\end{eqnarray}%zs3w

\begin{figure}
\includegraphics[angle=0,width=0.45\textwidth] {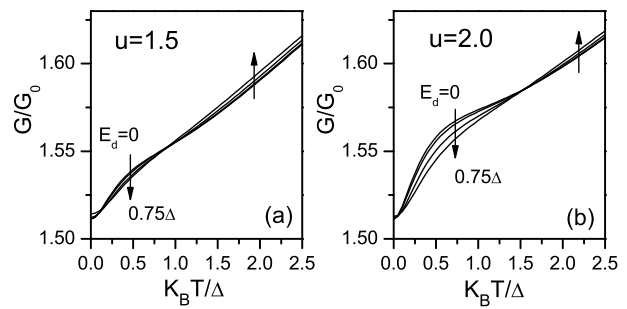}
\caption{\label{FIG:Rt} R versus temperature for $u=1.5$ and $u=2.0$
for different Ed}
\end{figure}

We investigate the effects on the conductance at finite temperature. The increasing temperature
suppress the Kondo resonance and enhance the conductance at $E_F$. The results are given in Fig. 2
for different parameters. The conductance increases with the increasing temperature, and then get
saturated. Increasing temperature enhances the conductance G, which is clear the evidence of Kondo
effects. The transport is ballistic and contributes to $2G_0$ at any temperature as no impurities
appear in a nanotube. When the magnetic impurity emerges, the electrons suffer scattering when
passing through the conductive channel. In Fig. 2(a), the Coulomb strength $u=1.5$, the
conductance changes smaller and gets saturated at lower value. The Coulomb strength $u=2.0$ in
Fig. 2(b). For different $E_d$, the conductance behaves differently. For small $E_d$, the
conductance starts at a higher value and changes relatively smaller with respect to the bigger
$E_d$.

%%%%%%%%%%%%%%%%%%%%%%%%%%%%%%%%%%%%%%%%%%
\subsection{Thermoelectric power}
Thermoelectric power (TEP) is an important transport coefficient, because it is related to the
energy derivative of electrical conductivity at the $E_F$. The contribution of magnetic impurities
to the low-temperature diffusion TEP $S$ is calculated by Mott relation:
\begin{eqnarray}
S(T)=-\frac{-\pi^2k_B^2}{3e}T\left(\frac{d\mbox{ln}\sigma(\varepsilon)}{d\varepsilon}\right)_{E_F}
\end{eqnarray}
It is valid in one-dimensional system by incorporating the Landauer
formula for the conductance.\cite{sivan}

\begin{figure}
\includegraphics[angle=0,width=0.45\textwidth] {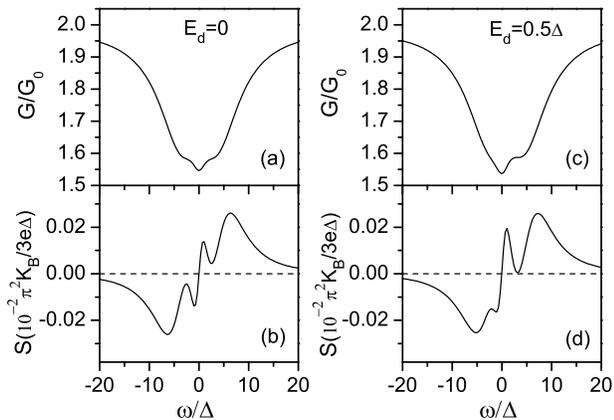}
\caption{\label{FIG:Se} TEP for $u=1.5$ and $u=2.0$ for different
$E_d$}
\end{figure}

TEP is sensitive to the position of Kondo resonance peak, which determine the sign of TEP. TEP of
an individual SWNT has been measured, and gate electric field dependent TEP modulation is found.
The effects of defects cause the valley of conductance, and then TEP will change the sigh when
changing gate voltage $V_g$. We give the results of G and  at $T_K=0.001\Delta$ for a (10, 10)
nanotube. But here the Kondo resonace make the structure different.  For the symmetric Anderson
model at low temperature, the TEP changes its sign, as shown in Fig. \ref{FIG:Se}.
%It is consistent with the observation in experiments.

\begin{figure}
\includegraphics[angle=0,width=0.45\textwidth] {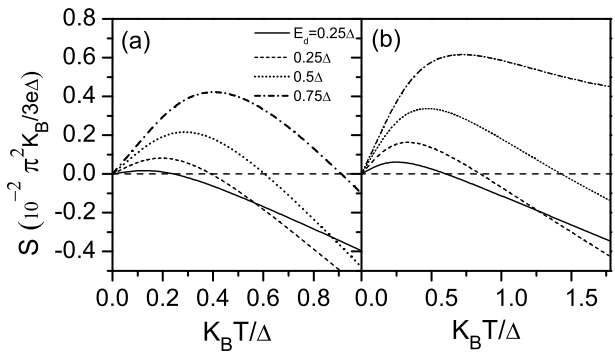}
\caption{\label{FIG:asymtep} TEP versus $K_BT$ for with $Ed=0,
0.25\Delta, 0.5\Delta, 0.25\Delta$ (a)$u=1.5$ and (b)$u=2.0$}
\end{figure}

For symmetric model, which keep the electron-hole symmetry, the TEP always equals to 0. But for
asymmetric model in Kondo regime, TEP increases and then decreases which can interpret the
observed phenomenon for the system of nanotubes with transition metal catalyst. Fig.
\ref{FIG:asymtep} gives the TEP as a function of $K_BT/\Delta$. (a) is for the case of u=1.5 and
(b) corresponds to the case of u=2.0. It is clear shown that the TEP gets a increase then decrease
process. It is the evidence of Kondo effects. Comparing the two figures, we find that for bigger
$E^0_d$ the TEP will need higher temperature to reach the peak value. It is the same for smaller
u.

\begin{figure}
\includegraphics[angle=0,width=0.30\textwidth] {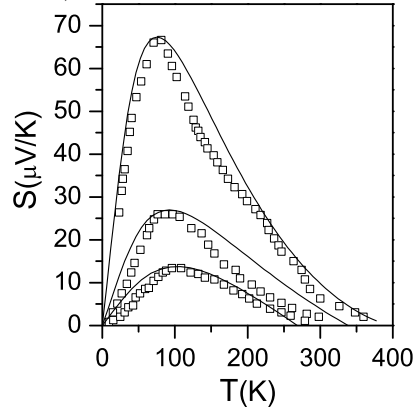}
\caption{\label{FIG:tepexp} TEP versus T in the absolute unit, the
dot is the experimental data for Ref.\ref{ff}}
\end{figure}
Since the u and $E^0_d$ relates to the Kondo temperature $T_k$, here we study the TEP in the scale
of $T_k$. At low bias, the $E_F$ can be assumed that it does not shift when the density of
impurities is small. Fig. 6 gives the results for different TEP versus $K_BT_k/\Delta$, $T_k$ is
given by:\cite{costi}
\begin{equation}
K_BT_k=U(\frac{\Delta}{2U})^{1/2}e^{\pi\varepsilon(\varepsilon+U)/2\Delta U}
\end{equation}%
We adjust the x scale according the $T_k$, and give results
consistent with the experimental measurements.

\section{Summary}
By introducing the normal self-energy contributed by the magnetic impurities in the single-walled
metallic carbon nanotubes, we have studied the electric transport properties at finite temperature
within the Anderson model and Landauer formula. The magnetic impurities induce Kondo resonance
near the $E_F$, and hence the conductance valley appeared. As the temperature increases, the Kondo
resonance and the conductance valley fade out for both the symmetric and asymmetric case. For
bigger radius nanotubes, the width of resonant peak and the conductance valley at low temperature
is suppressed. With the increasing temperature, the conductance of the nanotube increases in the
low temperature range.

Based on the Mott relation, the TEP of carbon nanotubes with magnetic impurities has been also
investigated. It changes the sign near the $E_F$ due to the Kondo resonance. At low temperature
TEP increases with the increasing temperature, and then it decreases after reaching a peak value.
The stronger Couloumb strength or the bigger $E_d$ is, the earlier the TEP reaches the peak value.
When scaled according to the Kondo temperature, the dependence of TEP on the temperature is found
to interpret the experiment well.

Our studies on the magnetic impurities in metallic carbon nanotubes gives Kondo characters of
electric transport properties in mesoscopic system. Further experiments would be expected to
explore the roles of magnetic impurities in the carbon nanotubes.

\section{Acknowledgements}
The authors gratefully acknowledge financial supports from NSF China (Grant No. 10574077),
the``863" Programme of China (No. 2006AA03Z0404) and the ``973" Programme of China (No.
2005CB623606) are gratefully acknowledged.

\bibliography{MagNT3p}
\end{document}